\documentclass[%twocolumn,%showpacs,preprintnumbers,
amsmath,%amssymb,aps,
prd,nofootinbib,floatfix,12pt,%preprint,
]{revtex4}

\def\MSbar{\overline{\rm MS}}
\def\gptwo{g^{\prime 2}}
\def\gpfour{g^{\prime 4}}
\def\lnbar{\overline{\ln}}
\newcommand\beq{\begin{eqnarray}}
\newcommand\eeq{\end{eqnarray}}
\newcommand\Tbar{\overline{T}}

\def\lsim{\mathrel{\rlap{\lower4pt\hbox{$\sim$}}
    \raise1pt\hbox{$<$}}}                % less than or approx. symbol
\def\gsim{\mathrel{\rlap{\lower4pt\hbox{$\sim$}}
    \raise1pt\hbox{$>$}}}            

\allowdisplaybreaks
\usepackage{graphicx}% Include figure files
\usepackage{setspace}
\usepackage{bm}% bold math

\begin{document}

\renewcommand{\theequation}{\arabic{section}.\arabic{equation}}
\renewcommand{\thefigure}{\arabic{section}.\arabic{figure}}
\renewcommand{\thetable}{\arabic{section}.\arabic{table}}

\title{\large \baselineskip=20pt 
Higgs boson mass in the Standard Model at two-loop order\\ and beyond}

\author{Stephen P.~Martin$^{1,2}$ and David G.~Robertson$^3$}
\affiliation{
{\it $^1$Department of Physics, Northern Illinois University, DeKalb IL 60115} \\
{\it $^2$Fermi National Accelerator Laboratory, P.O. Box 500, Batavia IL 60510}
\\
{\it $^3$Department of Physics, Otterbein University, Westerville OH 43081}}

\begin{abstract}\normalsize\baselineskip=18pt 
We calculate the mass of the Higgs boson in the Standard Model in terms 
of the underlying Lagrangian parameters at complete 2-loop order with 
leading 3-loop corrections. A computer program implementing the results 
is provided. The program also computes and minimizes the Standard Model effective potential in Landau gauge at 2-loop order with leading 3-loop corrections.
\end{abstract}

\maketitle

\vspace{-0.3in}

\tableofcontents

\baselineskip=17pt

%%%%%%%%%%%%%%%%%%%%%%%%%%%%%%%%%%%%%%%%%%%%%%%%%%%%%%%%%%%%%%%%
\section{Introduction\label{sec:intro}}
\setcounter{equation}{0}
\setcounter{figure}{0}
\setcounter{table}{0}
\setcounter{footnote}{1}

The Large Hadron Collider (LHC) has discovered \cite{LHCdiscovery} a 
Higgs scalar boson $h$ with mass $M_h$ near 125.5 GeV \cite{LHCmass} and 
properties consistent with the predictions of the minimal Standard 
Model. At the present time, there are no signals or hints of other new 
elementary particles. In the case of supersymmetry, the limits on 
strongly interacting superpartners are model dependent, but 
typically extend to over an order of magnitude above $M_h$. It is 
therefore quite possible, if not likely, that the Standard Model with a 
minimal Higgs sector exists as an effective theory below 1 TeV, with all 
other fundamental physics decoupled from it to a very good 
approximation. Within this model, precision calculations can help to 
relate observable quantities to the underlying Lagrangian parameters, as 
well as help to constrain new physics models, including those for which 
decoupling may not hold.

One such observable quantity is the physical mass $M_h$ itself. At tree 
level, $M_h$ is directly proportional to the square root of the 
Higgs field self 
interaction coupling, $\lambda$. One important question has to do
with the stability of the Standard Model vacuum
\cite{Sher:1988mj,Lindner:1988ww,Arnold:1991cv,Ford:1992mv,Casas:1994qy,Espinosa:1995se,Casas:1996aq,Isidori:2001bm,Espinosa:2007qp,ArkaniHamed:2008ym,Bezrukov:2009db,Ellis:2009tp}.
The observed value of $M_h$ is in the 
range that would apparently correspond to metastability of the 
vacuum \cite{EliasMiro:2011aa,Alekhin:2012py,Bezrukov:2012sa,Degrassi:2012ry,Buttazzo:2013uya},
assuming that there is no new physics
between the electroweak scale and the Planck scale. 
It is therefore important to pin down the relationship 
between $\lambda$ and $M_h$ as accurately as possible. Parametric 
uncertainties, notably the dependences on the top-quark mass and the QCD 
coupling, are not insignificant, and will likely 
remain so for some time. However, our attitude is that theoretical 
calculations should, to the extent possible, be pushed to the point that all 
limitations of our understanding can be reliably and unambiguously
blamed on experimental error.
 
The purpose of this paper is to present a full 2-loop calculation of the 
minimal Standard Model Higgs boson pole mass $M_h$, in terms of the 
$\MSbar$ Lagrangian parameters $v, \lambda, y_t, g, g', g_3$, with the leading
3-loop corrections in the limit $g_3, y_t \gg \lambda, g, g'$. The 
relations between these parameters and other observables, such as the 
physical masses of the top quark and the $Z$ and $W$ bosons, are left to 
separate calculations. The result for $M_h$ is probably too long to 
present as an analytical formula in print without forfeiting the 
goodwill of the reader, and in any case evaluation of it will 
necessarily rely on numerical work done by computer. We therefore 
present most of our results in the form of an electronic file, and as a 
public computer code. The computer code also performs the related 
task of minimizing the 2-loop effective potential \cite{FJJ} of the 
Standard Model with leading 3-loop corrections \cite{Martin:2013gka}, 
implementing the form of the minimization condition given recently in 
\cite{Martin:2014bca,Elias-Miro:2014pca}, which resummed Goldstone 
contributions to eliminate spurious imaginary parts and potentially 
infrared singular contributions.

Previous work on the 2-loop contributions to the relation between 
$\lambda$ and $M_h$ includes the QCD corrections 
\cite{Bezrukov:2012sa,Degrassi:2012ry}, which can be obtained from the 
2-loop QCD correction \cite{Kniehl:1994ph,Djouadi:1994gf} to the Higgs 
self-energy function. The non-QCD corrections have been obtained by 
\cite{Degrassi:2012ry} and \cite{Buttazzo:2013uya} but were given there 
only in the form of simple interpolating formulas.

%%%%%%%%%%%%%%%%%%%%%%%%%%%%%%%%%%%%%%%%%%%%%%%%%%%%%%%%%%%%%%%%
\section{Higgs pole mass at 2-loop order\label{sec:Mh}}
\setcounter{equation}{0}
\setcounter{figure}{0}
\setcounter{table}{0}
\setcounter{footnote}{1}

To fix our conventions and
notation, we write the Higgs kinetic and self-interaction 
Lagrangian as
\beq
{\cal L} =  -\partial^\mu \Phi^\dagger \partial_\mu \Phi
-\Lambda -m^2 \Phi^\dagger \Phi -\lambda (\Phi^\dagger \Phi)^2 ,
\eeq
where we use the metric with signature ($-$,$+$,$+$,$+$), 
and $m^2 < 0$, and the complex doublet Higgs field is
\beq
\Phi(x) = \begin{pmatrix}
\frac{1}{\sqrt{2}} [v + h(x) + i G^0(x)]
\\
G^+(x)
\end{pmatrix}.
\eeq
Here $v$ is the Higgs vacuum expectation value (VEV), which we take to 
be evaluated at the minimum of the effective potential evaluated at 
2-loop order with leading 3-loop corrections. 
This means that the sum of tadpole diagrams 
(including the tree-level one)
vanishes at that same order, and so need not be included. Because the 
Landau gauge is used for the evaluation of the effective 
potential in \cite{FJJ}-\cite{Elias-Miro:2014pca}, our calculation also 
is restricted to that gauge-fixing scheme.

The other relevant couplings in the theory are the top-quark Yukawa 
coupling $y_t$ and the $SU(3)_c \times SU(2)_L \times U(1)_Y$ gauge 
couplings $g_3$, $g$, $g'$. In principle, the bottom quark and other 
fermion Yukawa couplings can also be included, but they make only a very 
tiny difference even at 1-loop order, where their inclusion is 
straightforward (see below). All of the couplings $\lambda$, $m^2$, 
$y_t$, $g_3$, $g$, $g'$, and the VEV $v$, are running parameters in the 
$\MSbar$ scheme.

In order to obtain the Higgs boson physical mass $M_h$, 
we calculate the self-energy
function
\beq
\Pi(s) = \frac{1}{16 \pi^2} \Pi^{(1)}(s) 
\,+\, \frac{1}{(16 \pi^2)^2} \Pi^{(2)}(s) 
+ \ldots
\eeq
consisting of the sum of all 1-particle-irreducible 2-point Feynman 
diagrams, in the regulated theory in $d=4-2\epsilon$ dimensions. In 
this paper, factors of $1/(16 \pi^2)^\ell$ are extracted as a way of signifying 
the loop order $\ell$. Rather than including counterterm diagrams separately, 
we found it more efficient to do the calculation in terms of the bare 
quantities $\lambda_B$, $m^2_B$, $y_{tB}$, $g_{3B}$, $g_B$, $g'_B$, and 
VEV $v_B$, and then re-express 
the results in terms of the $\MSbar$ quantities. The complex pole 
squared mass is the solution of
\beq
M^2_h - i \Gamma_h M_h \>\equiv\> s_{\rm pole} \>=\>
m^2_B + 3 \lambda_B v^2_B + 
\frac{1}{16 \pi^2} \Pi^{(1)}(s_{\rm pole}) \,+\, 
\frac{1}{(16 \pi^2)^2} \Pi^{(2)}(s_{\rm pole}),
\label{eq:M2hbare}
\eeq
where 3-loop order effects are consistently neglected in this section. We then apply
the $\MSbar$ relations between bare and renormalized parameters:
\beq
\label{eq:v2B}
v^2_B &=& \mu^{-2\epsilon} v^2 \Bigl [
1 + \frac{1}{16\pi^2} \frac{c^\phi_{1,1}}{\epsilon} 
+ \frac{1}{(16\pi^2)^2} \Bigl ( \frac{c^\phi_{2,2}}{\epsilon^2}
+ \frac{c^\phi_{2,1}}{\epsilon} \Bigr ) + \ldots \Bigr ]
,
\\
\lambda_B &=& \mu^{2\epsilon} \Bigl [
\lambda + \frac{1}{16\pi^2} \frac{c^\lambda_{1,1}}{\epsilon} 
+ \frac{1}{(16\pi^2)^2} \Bigl ( \frac{c^\lambda_{2,2}}{\epsilon^2}
+ \frac{c^\lambda_{2,1}}{\epsilon} \Bigr ) + \ldots \Bigr ]
,
\\
m^2_B &=& 
m^2 + \frac{1}{16\pi^2} \frac{c^{m^2}_{1,1}}{\epsilon} 
+ \frac{1}{(16\pi^2)^2} \Bigl ( \frac{c^{m^2}_{2,2}}{\epsilon^2}
+ \frac{c^{m^2}_{2,1}}{\epsilon} \Bigr ) + \ldots
,
\\
y_{tB} &=& \mu^{\epsilon} \Bigl [
y_t + \frac{1}{16\pi^2} \frac{c^{y_t}_{1,1}}{\epsilon} 
+ \ldots \Bigr ]
,
\\
g_{B} &=& \mu^{\epsilon} \Bigl [
g + \frac{1}{16\pi^2} \frac{c^{g}_{1,1}}{\epsilon} 
+ \ldots \Bigr ]
,
\\
g'_{B} &=& \mu^{\epsilon} \Bigl [
g' + \frac{1}{16\pi^2} \frac{c^{g'}_{1,1}}{\epsilon} 
+ \ldots \Bigr ]
,
\\
g_{3B} &=& \mu^{\epsilon} \left [
g_3 + \ldots \right ]
\label{eq:g3B}
,
\eeq
to obtain $s_{\rm pole}$ in terms of the renormalized parameters. 
Here $\mu$ is the regularization scale,
related to the $\MSbar$ renormalization scale $Q$ by
\beq
Q^2 = 4\pi e^{-\gamma_E} \mu^2,
\eeq
where $\gamma_E = 0.5772\ldots$ is the Euler-Mascheroni constant, and
the counterterm coefficients are, to the orders required for this paper:
\beq
c_{1,1}^\phi &=& -3 y_t^2 + \frac{9}{4} g^2 + \frac{3}{4} \gptwo
,
\\
c_{2,2}^\phi &=& 
12 g_3^2 y_t^2 
-\frac{9}{4} y_t^4 
-\frac{27}{8} y_t^2 g^2 
- \frac{1}{8} y_t^2 \gptwo
- \frac{33}{32} g^4 
+ \frac{27}{16} g^2 \gptwo
+ \frac{91}{32} \gpfour
,
\\
c_{2,1}^\phi &=& 
-10 g_3^2 y_t^2 
+\frac{27}{8} y_t^4 
-\frac{45}{16} y_t^2 g^2 
- \frac{85}{48} y_t^2 \gptwo
+ \frac{271}{64} g^4 
- \frac{9}{32} g^2 \gptwo
- \frac{431}{192} \gpfour
- 3 \lambda^2
,
\\
c_{1,1}^\lambda &=& 
-3 y_t^4 
+ 6 \lambda y_t^2 
+ 12 \lambda^2 
- \frac{9}{2} \lambda g^2 
- \frac{3}{2} \lambda \gptwo
+ \frac{9}{16} g^4 
+ \frac{3}{8} g^2 \gptwo 
+ \frac{3}{16} \gpfour
,
\\
c_{2,2}^\lambda &=& 
24 g_3^2 y_t^4 
- 24 g_3^2 y_t^2 \lambda
-\frac{45}{2} y_t^6 
+ \frac{27}{2} y_t^4 g^2 
+ \frac{13}{2} y_t^4 \gptwo
-\frac{9}{2} y_t^4 \lambda 
+ 108 y_t^2 \lambda^2 
- \frac{135}{4} y_t^2 \lambda g^2
\phantom{xxx}
\nonumber \\ &&
-\frac{53}{4} y_t^2 \lambda \gptwo
+ \frac{27}{16} y_t^2 g^4 
+ \frac{9}{8} y_t^2 g^2 \gptwo 
+ \frac{9}{16} y_t^2 \gpfour
+ 144 \lambda^3
- 81 \lambda^2 g^2
- 27 \lambda^2 \gptwo
\nonumber \\ &&
+ 24 \lambda g^4
+ \frac{45}{4} \lambda g^2 \gptwo
- \frac{7}{4} \lambda \gpfour
- \frac{195}{64} g^6
- \frac{119}{64} g^4 \gptwo
+ \frac{37}{64} g^2 \gpfour
+ \frac{73}{64} g^{\prime 6}
,
\\
c_{2,1}^\lambda &=& 
-8 g_3^2 y_t^4 
+ 20 g_3^2 y_t^2 \lambda
+\frac{15}{2} y_t^6 
- \frac{2}{3} y_t^4 \gptwo
-\frac{3}{4} y_t^4 \lambda 
-36 y_t^2 \lambda^2 
+ \frac{45}{8} y_t^2 \lambda g^2
\nonumber \\ &&
+\frac{85}{24} y_t^2 \lambda \gptwo
- \frac{9}{16} y_t^2 g^4 
+ \frac{21}{8} y_t^2 g^2 \gptwo 
- \frac{19}{16} y_t^2 \gpfour
-78 \lambda^3
+ 27 \lambda^2 g^2
+ 9 \lambda^2 \gptwo
\nonumber \\ &&
- \frac{73}{32} \lambda g^4
+ \frac{39}{16} \lambda g^2 \gptwo
+ \frac{629}{96} \lambda \gpfour
+ \frac{305}{64} g^6
- \frac{289}{192} g^4 \gptwo
- \frac{559}{192} g^2 \gpfour
- \frac{379}{192} g^{\prime 6}
,
\\
c_{1,1}^{m^2} &=& m^2 \Bigl [
3 y_t^2 + 6 \lambda - \frac{9}{4} g^2 - \frac{3}{4} \gptwo
\Bigr ] ,
\\
c_{2,2}^{m^2} &=& m^2 \Bigl [
-12 g_3^2 y_t^2 
+ \frac{9}{4} y_t^4 
+ 36 y_t^2 \lambda
- \frac{81}{8} y_t^2 g^2
- \frac{35}{8} y_t^2 \gptwo
+ 54 \lambda^2
- 27 \lambda g^2
- 9 \lambda \gptwo
\nonumber \\ &&
+ \frac{249}{32} g^4 
+\frac{45}{16} g^2 \gptwo
- \frac{55}{32} \gpfour
\Bigr ] ,
\\
c_{2,1}^{m^2} &=& m^2 \Bigl [
10 g_3^2 y_t^2 
- \frac{27}{8} y_t^4 
-18 y_t^2 \lambda
+ \frac{45}{16} y_t^2 g^2
+ \frac{85}{48} y_t^2 \gptwo
- 15 \lambda^2
+ 18 \lambda g^2
+ 6 \lambda \gptwo
\nonumber \\ &&
- \frac{145}{64} g^4 
+ \frac{15}{32} g^2 \gptwo
+ \frac{557}{192} \gpfour
\Bigr ] ,
\\
c_{1,1}^{y_t} &=& y_t \Bigl [ 
-4 g_3^2 + \frac{9}{4} y_t^2
- \frac{9}{8} g^2 - \frac{17}{24} \gptwo \Bigr ] ,
\\
c_{1,1}^g &=& -19 g^3/12 ,
\\
c_{1,1}^{g'} &=& 41 g^{\prime 3}/12 . 
\label{eq:cgp1}
\eeq
These counterterm coefficients can be obtained from the
2-loop beta functions and anomalous dimension given in
refs.~\cite{MVI,MVII,Jack:1984vj,MVIII}, \cite{FJJ}; see for example
the discussion in eqs.~(4.5)-(4.14) of ref.~\cite{Martin:2013gka} 
which uses the same notations and conventions as the present paper.

\begin{figure}[t]
\includegraphics[width=0.86\linewidth,angle=0]{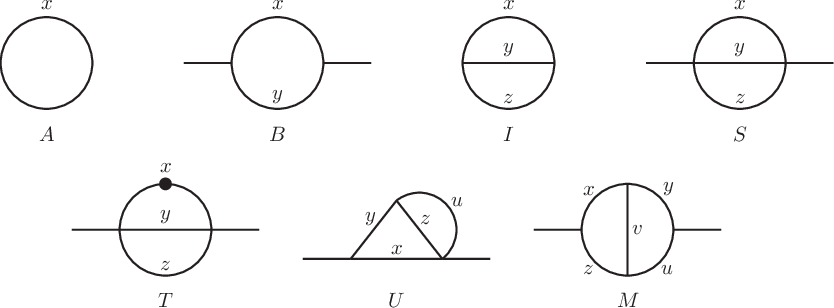}
\vspace{-0.25cm}
\caption{\label{fig:basis}
Topologies for the one- and two-loop vacuum and
self-energy scalar basis integrals used in this paper and defined in 
refs.~\cite{Martin:2003qz,TSIL}. The dot in the $T$ topology stands for
a derivative with respect to the squared mass $x$.}
\end{figure}

The 1-loop and 2-loop integrals are reduced, using the Tarasov algorithm
\cite{Tarasov:1997kx} implemented in the program TARCER \cite{Mertig:1998vk},
to a set of Euclidean $d$-dimensional scalar basis integrals 
with topologies 
illustrated in Figure \ref{fig:basis} and
defined in our notation in 
refs.~\cite{Martin:2003qz,TSIL}. The 1-loop integrals are
\beq
{\bf A}(x),\> {\bf B}(x,y),
\label{eq:boldAB}
\eeq
and the 2-loop integrals are
\beq
{\bf I}(x,y,z),\> {\bf S}(x,y,z),\> {\bf T}(x,y,z),\> {\bf U}(x,y,z,u),\> 
{\bf M}(x,y,z,u,v),
\label{eq:boldISTUM}
\eeq
where the arguments are bare squared masses. The integrals 
${\bf B}, {\bf S}, {\bf T}, {\bf U}$, and ${\bf M}$ also each have
an implicit dependence on the external momentum invariant $s = -p^2$.
The integrals have invariances under interchanges of squared mass arguments
that are obvious from the figures.
%%%%%%%%%%%%%%%%%%%%%%%%%%%%%%%%%%%%%%%%%%%%%%%%%%%%%%%%%%%%%%

In terms of bare quantities, the propagators in the self-energy integrals
depend on the squared masses of the neutral and charged Goldstone bosons,
the Higgs boson, the top quark, and the $W$ and $Z$ bosons:
\beq
G_B &=& m^2_B + \lambda_B v_B^2 ,
\\
H_B &=& m^2_B + 3 \lambda_B v_B^2,
\\
t_B&=& y_{tB}^2 v_B^2/2,
\\
W_B&=& g_B^2 v_B^2/4,
\\
Z_B&=& (g_B^2 + g_B^{\prime 2}) v_B^2/4,
\eeq
with a massless photon and ghosts. We then perform an expansion using
eqs.~(\ref{eq:v2B})-(\ref{eq:cgp1}), to write
these quantities in terms of the corresponding $\MSbar$ squared masses.
For the 2-loop integrals, this merely requires replacing the
bare squared mass arguments by their $\MSbar$ counterparts, because
the difference is of 3-loop order. For the 1-loop integrals, the 
functions ${\bf A}$ and ${\bf B}$ are expanded to first order
around the $\MSbar$ squared-mass arguments 
\beq
G &=& m^2 + \lambda v^2 ,
\\
H &=& m^2 + 3 \lambda v^2,
\\
t&=& y_{t}^2 v^2/2,
\\
W&=& g^2 v^2/4,
\\
Z&=& (g^2 + g^{\prime 2}) v^2/4,
\eeq
using 
\beq
{\bf A}(X) &=& {\bf A}(x) 
+ (X - x) \frac{\partial}{\partial x}{\bf A}(x) + \ldots
,
\label{eq:Aboldexp}
\\
{\bf B}(X, Y) &=& {\bf B}(x,y) + 
(X - x) \frac{\partial}{\partial x} {\bf B}(x,y) + 
(Y - y) \frac{\partial}{\partial y} {\bf B}(x,y) + \ldots
,
\label{eq:Bboldexp}
\eeq
where the derivatives are given in the Appendix.
As a further refinement, the parameter $m^2$ is eliminated using
the minimization condition of the Landau gauge effective potential,
which takes the form
\beq
G \>=\> m^2 + \lambda v^2 &=&
-\frac{1}{16 \pi^2} \widehat{\Delta}_1
-\frac{1}{(16 \pi^2)^2} \widehat{\Delta}_2
-\ldots ,
\label{eq:resummedmincon}
\eeq
given in eqs.~(4.18)-(4.21) of ref.~\cite{Martin:2014bca} 
(with equivalent results in \cite{Elias-Miro:2014pca}).
Here the quantities $\widehat{\Delta}_1$ and $\widehat{\Delta}_2$
depend on $t$, $W$, $Z$, and
\beq
h = 2 \lambda v^2,
\eeq
but not on $G$ or $H$ or $m^2$. 
The 1-loop integrals
involving $G$ as an argument are expanded using 
eqs.~(\ref{eq:AG})-(\ref{eq:BepsGx}) of the Appendix, 
while those involving $H$ as
an argument are expanded using 
eqs.~(\ref{eq:Aboldexp}), (\ref{eq:Bboldexp}) again.

The loop integrals are then rewritten in terms of the basis of
$\epsilon$-independent integrals
\beq
&&
A(x),\>B(x,y),\>I(x,y,z),\>S(x,y,z),\>T(x,y,z),\>\Tbar(0,x,y),\>
\nonumber \\ && 
U(x,y,z,u),\> M(x,y,z,u,v)
\label{eq:defABISTUM}
\eeq
which are obtained from the corresponding integrals in eqs.~(\ref{eq:boldAB}), 
(\ref{eq:boldISTUM}) by subtracting
appropriate sub-divergences and taking the limit $\epsilon \rightarrow 0$.
Here $\Tbar(x,y,z) \equiv T(x,y,z) + B(y,z) \lnbar(x) $ with 
\beq
\lnbar(x) &\equiv& \ln(x/Q^2).
\eeq
The reason for the definition of the function $\Tbar(x,y,z)$
is that it is well defined as $x \rightarrow 0$,
while $T(x,y,z)$ diverges in that limit.
For the precise definitions of the integrals 
in eq.~(\ref{eq:defABISTUM}), see section 2 of \cite{TSIL}.
These integrals also have an implicit dependence on the common 
external momentum invariant $s$ and on the $\MSbar$ renormalization scale $Q$.
In the resulting expression on the right-hand side of eq.~(\ref{eq:M2hbare}), 
there are terms proportional to
$s_{\rm pole}/\epsilon$ and $s_{\rm pole}/\epsilon^2$, corresponding 
to the Higgs wavefunction renormalization.  
These are moved to the left-hand side to allow $s_{\rm pole}$ to be solved for.
Finally, the 
regulator is removed by taking the limit $\epsilon \rightarrow 0$. 

The result for the Higgs squared pole mass is thus obtained in the form:
\beq
M^2_h - i \Gamma_h M_h &=& 2 \lambda v^2 
+ \frac{1}{16 \pi^2} \Delta_{M^2_h}^{(1)}
+ \frac{1}{(16 \pi^2)^2} \left [
\Delta_{M^2_h}^{(2),{\rm QCD}} + \Delta_{M^2_h}^{(2),{\rm non-QCD}} \right ],
\label{eq:M2hpole}
\eeq
where the right-hand side is a function of $v, \lambda, y_t, g, g', g_3,Q$,
with propagator
masses expressed as the combinations $h, t, W, Z$, and $0$. 
Working to 2-loop order with bottom, tau, and charm Yukawa couplings 
neglected, we can treat $s_{\rm pole}$ as real where it appears as
the (implicit) argument of the basis integral functions, and so replace it by 
$M_h^2$. This is because the imaginary part of $s_{\rm pole}$
is already of 2-loop order, 
and so the effect of including it would make a difference 
of 3-loop electroweak order in the pole mass. 
If the lighter fermions
are included in the 1-loop self-energy (see below), then there 
is a 1-loop imaginary part to the complex pole squared mass, but 
it is numerically smaller than a typical 3-loop order contribution due to
the small Yukawa couplings of $b$, $\tau$, $c$,
so that it can still be safely and consistently neglected. This feature is 
of course related to the very narrow Higgs width in the Standard Model.
For simplicity, we will therefore write $s=M_h^2$ below.

The complete lists of
1-loop and 2-loop basis integrals appearing on the right-hand side are
\beq
I^{(1)} &=& \bigl \{
B(t,t),\> B(h,h),\> B(W,W),\> B(Z,Z),\>
A(t),\>  A(h),\> A(W),\> A(Z)
\bigr \}
\eeq
and
%% 67 of these
\beq
I^{(2)} &=& \bigl \{
M(h,h,h,h,h),\> U(h,h,h,h),\> S(h,h,h),\> 
M(h,Z,h,Z,Z),\> U(h,h,Z,Z),\>
\nonumber \\ &&
M(W,W,W,W,h),\> U(W,W,W,h),\> S(h,W,W),\> T(W,W,h),\> 
\nonumber \\ &&
M(Z,Z,Z,Z,h),\> U(Z,Z,Z,h),\> S(h,Z,Z),\> T(Z,Z,h),\> 
\nonumber \\ &&
M(W,W,W,W,Z),\> U(W,W,W,Z),\> S(W,W,Z),\> T(W,W,Z),\> T(Z,W,W),\>
\nonumber \\ &&
M(W,Z,W,Z,W),\> U(Z,Z,W,W),\> M(h,W,h,W,W),\> U(h,h,W,W),\>
\nonumber \\ &&
M(t,t,t,t,Z),\> U(t,t,t,Z),\> S(t,t,Z),\> T(t,t,Z),\> T(Z,t,t),\> 
\nonumber \\ &&
M(t,t,t,t,h),\> U(t,t,t,h),\> S(h,t,t),\> T(t,t,h),\> 
%T(h,t,t),\> 
\nonumber \\ &&
M(t,Z,t,Z,t),\> U(Z,Z,t,t),\>  M(t,h,t,h,t),\> U(h,h,t,t),\>
\nonumber \\ &&
M(t,W,t,W,0),\> U(W,W,0,t),\> U(t,t,0,W),\> S(0,t,W),\> T(W,0,t),\> T(t,0,W),\> 
\nonumber \\ &&
M(t,t,t,t,0),\> T(t,0,t),\> \Tbar (0,t,t),\>
\nonumber \\ &&
M(W,W,W,W,0),\> T(W,0,W),\> \Tbar(0,W,W),\> 
U(W,W,0,0),\>  S(0,0,W),\> 
\nonumber \\ &&
T(W,0,0),\> 
U(Z,Z,0,0),\>  
S(0,0,Z),\> 
T(Z,0,0),\> 
I(h,h,h),\> 
I(t,t,Z),\> 
\nonumber \\ &&
I(h,t,t),\> 
I(W,W,Z),\> 
I(h,W,W),\> 
I(h,Z,Z),\> 
I(0,t,W),\> 
I(0,h,W),\> 
\nonumber \\ &&
I(0,h,Z),\> 
I(0,W,Z),\> 
I(0,0,W),\>
I(0,0,Z),\> 
I(0,0,h),\> 
I(0,0,t) 
\bigr \} .
\label{eq:I2list}
\eeq
In each of the $B$, $S$, $T$, $\overline T$, $U$, and $M$ integrals, the external
momentum invariant is taken to be the real pole squared mass, $s = M_h^2$, 
as discussed above. Then eq.~(\ref{eq:M2hpole}) can be solved numerically, by iteration.

The explicit results for the 1-loop part and the 2-loop QCD part of the 
Higgs pole squared mass corrections are:
\beq
\Delta_{M^2_h}^{(1)} &=& 
3 y_t^2 (4 t - s) B(t,t) 
- 18 \lambda^2 v^2 B(h,h)
\nonumber \\ &&
+ \frac{1}{2}(g^2 + \gptwo) \left [ (s - 3 Z - s^2/4Z) B(Z,Z) - s A(Z)/2Z + 2Z \right]
\nonumber \\ &&
+ g^2 \left [ (s - 3 W - s^2/4W) B(W,W) - s A(W)/2W + 2W \right] ,
\label{eq:Delta1M2h}
\\
\Delta_{M^2_h}^{(2),{\rm QCD}} &=& 
g_3^2 y_t^2 \Bigl [
8 (4t-s) (s-2t) M(t,t,t,t,0)
+ (36 s - 168 t) T(t,0,t)
\nonumber \\ &&
+ 16 (s-4t) \Tbar (0,t,t)
+ 14 s B(t,t)^2 
+ (-176 + 36 s/t) A(t) B(t,t)
\nonumber \\ &&
+ (80 t - 36 s) B(t,t)
- 28 A(t)^2/t\, + 80 t - 17 s
\Bigr ] .
\label{eq:Delta2M2hQCD}
\eeq
In eq.~(\ref{eq:Delta1M2h}), a term ${3\lambda} (s^2 - h^2) B(0,0)/h$ 
coming from loops involving Goldstone bosons and longitudinal vector bosons 
has been moved into the
2-loop order non-QCD 
part discussed below, by iterating using $s = h + \Delta_{M^2_h}^{(1)}/16\pi^2$.
There, it cancels against other terms, and the full 2-loop 
result does not depend on $B(0,0)$.
This is as expected, because a term with $B(0,0)$ coming from loops involving Goldstone bosons and longitudinal vector bosons
would imply an imaginary part to the
pole squared mass that does not correspond to any physical decay of the Higgs boson.
One-loop contributions $B(0,Z)$ and $B(0,W)$ from individual Feynman diagrams 
involving Goldstone bosons and longitudinal vector bosons also
cancel as expected, even without iteration in $s$.

For the remaining, non-QCD, 2-loop contributions, 
there are a large number of terms, and some of them
are a bit complicated, so 
that the length of the result may exceed the threshold of impoliteness,
and we decline to present them explicitly in print. 
The result has the form:
\beq
\Delta_{M^2_h}^{(2),{\rm non-QCD}} &=& 
\sum_i c^{(2)}_i I_i^{(2)}
+ \sum_{j\leq k} c_{j,k}^{(1,1)} I_j^{(1)} I_k^{(1)}
+ \sum_j c^{(1)}_j I_j^{(1)}
+ c^{(0)} .
\label{eq:Delta2M2hnonQCD}
\eeq
The coefficients $c^{(2)}_i$ and $c_{j,k}^{(1,1)}$ and $c^{(1)}_j$
and $c^{(0)}$ are available in electronic form in a file called
{\tt coefficients.txt}.
They are also implemented in a public computer code written in C, 
described below. These electronic files are available from the authors'
web pages \cite{webpages}, 
and {\tt coefficients.txt} is also included as an ancillary file with the
arXiv source for this article.
In these coefficients, we replaced $s$ by its tree-level
approximation $2\lambda v^2$ wherever it appears explicitly
(but not where it appears as the implicit argument of the basis 
functions). This enforces the cancellations between Goldstone and longitudinal
vector boson contributions, avoiding spurious imaginary contributions to the pole squared mass that do not correspond to physical decay modes of the Higgs boson.
Therefore each coefficient is a sum of ratios of polynomials in 
$\lambda, y_t, g, g'$, 
multiplied by the appropriate power of $v$.
The impact incurred by doing these substitutions for $s$ is 
of 3-loop order without involving QCD, and so is 
beyond the order of our calculations in this paper, including the QCD part of the leading 3-loop corrections discussed in the next section. 

The expression of the result in terms of the 
basis integrals is not unique, because there are identities 
between different basis integrals that hold when the squared mass arguments 
are not generic. These identities include 
eqs.~(A.14), (A.15), and (A.17)-(A.20) in ref.~\cite{Martin:2003it},
and eqs.~(\ref{eq:I00x})-(\ref{eq:U0x00}) in the Appendix 
of the present paper. We also used the threshold integral relations 
(\ref{eq:B0h}) and (\ref{eq:Th00}) 
in the Appendix to simplify the 2-loop order non-QCD part.

There are several quite non-trivial checks on the calculation. First, we 
checked that all single and double poles in $\epsilon$ cancel in 
$M^2_h$. This relies on agreement between the counter-term poles 
$c^X_{\ell,n}$ (for $X = v, \lambda, m^2, y_t, g, g'$) as extracted 
from the Higgs anomalous dimension and the beta functions in the 
literature, and the divergent parts of the loop integrations performed 
independently here. Second, we checked that logarithms of $G$ cancel, 
avoiding any spurious imaginary parts that would occur if the 
renormalization scale were chosen so that $G<0$, or spurious
divergences that would occur if $G=0$. Third, we observed 
cancellation between the parts of loop integral functions involving 
Landau gauge vector propagators with poles at squared mass equal to 0 
and the corresponding Goldstone propagators, once the latter were 
expanded using eq.~(\ref{eq:resummedmincon}). This is important in 
verifying the absence of spurious absorptive (imaginary) parts of the 
self-energy evaluated on-shell. Fourth, we noted that the imaginary
part $-i \Gamma_h M_h$ 
of eq.~(\ref{eq:M2hpole}) comes entirely from the contributions of
the six basis integrals
$U(W,W,0,0), S(0,0,W), T(W,0,0)$ and $U(Z,Z,0,0), S(0,0,Z), T(Z,0,0)$,
corresponding to the 3-body decays $\Gamma(h \rightarrow Wf\overline f')$
and $\Gamma(h \rightarrow Z f \overline f)$. We checked numerically to
very high precision that these imaginary contributions, when computed with
$s=h$, agree with the tree-level prediction for the
3-body widths found in eqs.~(8a)-(10)
of ref.~\cite{Keung:1984hn}. Fifth, we checked that although some of
the individual 2-loop coefficients in eq.~(\ref{eq:Delta2M2hnonQCD})
are singular in the formal limits 
$g,g'\rightarrow 0$ or $\lambda \rightarrow 0$, the whole expression is 
well-behaved in those limits, thanks to relations between 
different basis integrals when squared mass arguments are small.
Finally, we checked that the result for 
$M_h^2$ is renormalization group scale invariant 
through terms of 2-loop order. This is in principle 
equivalent to the first check mentioned, but in practice it tests the 
validity of various intermediate steps. It takes the form:
\beq
0 \> = \> Q\frac{d}{dQ} M^2_h &=& 
\left [
Q \frac{\partial}{\partial Q}
- \gamma_\phi v \frac{\partial}{\partial v}
+ \sum_X \beta_{X} \frac{\partial}{\partial X} 
\right ] M^2_h ,
\label{eq:RGinvariance}
\eeq
where $X = \{\lambda, y_t, g, g', g_3\}$, and $\gamma_\phi$ is the anomalous
dimension of the Higgs field. This check makes use of the derivatives of 
basis integrals with respect to the implicit argument $Q$, provided in 
eqs.~(4.7)-(4.13) of ref.~\cite{Martin:2003qz}, and 
on eqs.~(\ref{eq:dAdx}), (\ref{eq:dBxydx}) in the Appendix 
of the present paper. It also makes use of the
$\MSbar$ beta functions and Higgs anomalous dimension 
given in refs.~\cite{MVI,MVII,Jack:1984vj,MVIII}, \cite{FJJ}, 
\cite{Chetyrkin:2013wya,Bednyakov:2013eba}.

Although the lighter quarks and leptons have been neglected above due to their
very small Yukawa couplings, it is easy enough to include them in the leading 
approximation:
\beq
\Delta^{(1),b,\tau,c,\ldots}_{M^2_h} &=&
-[3 y_b^2 + y_\tau^2 + 3 y_c^2 + \ldots] B(0,0) M^2_h,
\label{eq:btauc}
\eeq 
Here we have taken $s = M_h^2$ and dropped the $y_f^4$ contributions and replaced the 
masses in light fermion propagators by 0. 
In that limit, we can also take
\beq
B(0,0) &=& 2 - \ln(M^2_h/Q^2) + i \pi.
\eeq
The numerical impact on the real pole mass $M_h$ from eq.~(\ref{eq:btauc})
is seen to be of order 1 MeV.
By comparing 
the imaginary part of the pole squared mass, $M^2_h - i \Gamma_h M_h$,
to the contribution of eq.~(\ref{eq:btauc}), multiplied by the loop factor
$1/16\pi^2$, we also obtain
the well-known result
\beq
\Gamma(h \rightarrow f \overline f) &=& \frac{N_c y_f^2 }{16 \pi} M_h .
\eeq
However, there are certainly better ways of obtaining the precise 
Higgs decay widths in the Standard Model; see for example ref.~\cite{HDECAY}
and references therein.

%%%%%%%%%%%%%%%%%%%%%%%%%%%%%%%%%%%%%%%%%%%%%%%%%%%%%%%%%%%%%%%%
\section{Leading three-loop corrections to the Higgs mass\label{sec:Mhalt}}
\setcounter{equation}{0}
\setcounter{figure}{0}
\setcounter{table}{0}
\setcounter{footnote}{1}

In this section, we find the leading 3-loop contributions to the 
Higgs pole squared mass in the effective potential approximation, 
based on the formal limit in which the top-quark squared mass is taken to be 
much larger than the squared masses of $h$, $Z$, and $W$.
In that limit, the Higgs self-energy function at leading order in
$y_t$ and $g_3$ can be approximated by taking $s=0$,
and is proportional to the second derivative of the renormalized 
effective potential
with respect to the Higgs field. Taking into account also the change in the 
minimization condition of the effective potential, we have a contribution (see for example section VI of ref.~\cite{Martin:2013gka}):
\beq
\delta M_h^2 &=& 
\left [
\frac{\partial^2}{\partial v^2}
-\frac{1}{v} \frac{\partial}{\partial v} \right ] \delta V_{\rm eff}.
\eeq
Using the leading 3-loop effective potential 
of ref.~\cite{Martin:2013gka}, with resummed Goldstone boson 
contributions to eliminate
spurious imaginary parts and infrared singular 
contributions \cite{Martin:2014bca,Elias-Miro:2014pca}, we obtain
the 3-loop contribution to be added to eq.~(\ref{eq:M2hpole}):
\beq
\Delta M_h^2 &=& \frac{1}{(16\pi^2)^3} \left [
\Delta_{M_h^2}^{(3),\mbox{leading QCD}} + \Delta_{M_h^2}^{(3),\mbox{leading non-QCD}}
\right ]
\label{eq:M2hpole3a}
\eeq
where 
\beq
\Delta_{M_h^2}^{(3),\mbox{leading QCD}} &=& 
g_3^4 y_t^2 t \left [248.122  + 839.197 \lnbar(t) + 160 \lnbar^2(t) -736 \lnbar^3(t) \right ]
\nonumber 
\\ &&
+ g_3^2 y_t^4 t \left [2764.365 + 1283.716 \lnbar(t) -360 \lnbar^2(t) + 240 \lnbar^3(t) \right ]
,
\phantom{xxxx}
\label{eq:M2hpole3b}
\eeq
\beq
\Delta_{M_h^2}^{(3),\mbox{leading non-QCD}} &=&
y_t^6 t \Bigl [
-3433.724
+ 36 \lnbar(h)
- 2437.511 \lnbar(t) 
+ 756 \lnbar(h) \lnbar(t) 
\nonumber \\ &&
- \frac{513}{2} \lnbar^2(t) 
+ 324 \lnbar(h) \lnbar^2(t) 
- 333 \lnbar^3(t) 
\Bigr ]
.
\label{eq:M2hpole3c}
\eeq
The analytical forms of the decimal coefficients are:
\beq
248.122 &\approx& -\frac{3776}{9} + 320 \zeta(3) + \frac{704\pi^4}{135}
+ \frac{256}{9} \ln^2(2) [\pi^2 - \ln^2(2)] -  \frac{2048}{3} {\rm Li}_4 (1/2)
,
\\ 
839.197 &\approx& 128 \zeta(3) + 2056/3 
,
\\
2764.365 &\approx& \frac{760}{3} - \frac{16\pi^2}{3} + 576 \zeta(3) 
+ \frac{496\pi^4}{15}
+ \frac{512}{3} \ln^2(2) [\pi^2 - \ln^2(2)]
- 4096 {\rm Li}_4(1/2),
\phantom{xxxx}
\\
1283.716 &\approx& -344 + 48 \pi^2 + 960 \zeta(3)
,
\\
-3433.724  &\approx& -673 - \frac{17 \pi^2}{2} - 1962 \zeta(3) - \frac{88\pi^4}{15}
- 32 \ln^2(2) [\pi^2 - \ln^2(2)]
+ 768 {\rm Li}_4 (1/2)
,\phantom{xxxx}
\\
-2437.511 &\approx& -\frac{3759}{2} - 39 \pi^2 - 144 \zeta(3)
.
\eeq
The 3-loop approximate formulas just described may be subject to significant 
corrections, because $s/t \approx 0.59$ is not a very small expansion parameter. 
However, experience shows that in such small-$s$ expansions of loop integrals
the coefficients of $s/t$ are typically also less than 1, 
so that the 3-loop approximation above might be expected to provide the bulk of the 
effect. For example, the small $s$-expansions of the 1-loop and 2-loop basis functions
involved in the contributions from the top quark and gluons are \cite{Martin:2003qz}:
\beq
B(t,t) &=& - \lnbar(t) + \frac{s}{6t} + \ldots
\label{eq:expBs}
\\
T(t,0,t) &=& \frac{1}{2} [\lnbar(t)-1]^2 + \frac{s}{4t} + \ldots
\\
\overline T (0,t,t) &=& -\frac{1}{2} [3 + 2 \lnbar(t) + \lnbar^2(t)] +
\frac{s}{36 t}[6 \lnbar (t) + 1] + \ldots
\\
M(t,t,t,t,0) &=& \frac{1}{t} + \frac{13s}{72t^2} + \ldots
\eeq
As noted in the discussion surrounding eqs.~(6.21)-(6.28) of 
ref.~\cite{Martin:2013gka}, the relatively small coefficient 248.122 
of the $g_3^4 y_t^2 t$ term 
independent of $\lnbar(t)$ in eq.~(\ref{eq:M2hpole3b}) 
is the result of a remarkable accidental near-cancellation. 
Because of this, the $g_3^2 y_t^4 t$ and $y_t^6 t $ contributions are actually 
numerically more important than the $g_3^4 y_t^2 t$ contribution.

Because the full $s$ dependence of the 2-loop QCD part was retained above,
the QCD part of the 3-loop contribution found in the effective potential 
approximation can simply be added in. 
As a check, we have verified the renormalization group invariance of the combined
full 2-loop plus leading 3-loop QCD result from eqs.~(\ref{eq:M2hpole}) and
(\ref{eq:Delta1M2h})-(\ref{eq:Delta2M2hnonQCD}) and
eqs.~(\ref{eq:M2hpole3a})-(\ref{eq:M2hpole3b}). 
This check consists of evaluating eq.~(\ref{eq:RGinvariance})
including all terms of 2-loop order
and the terms of 3-loop order that involve $g_3$ and are not suppressed by 
$\lambda$, $g$, or $g'$. The check again makes use of the
$\MSbar$ beta functions and Higgs anomalous dimension 
given in refs.~\cite{MVI,MVII,Jack:1984vj,MVIII}, \cite{FJJ}, 
\cite{Chetyrkin:2013wya,Bednyakov:2013eba}, as well as 
eqs.~(4.7)-(4.13) of ref.~\cite{Martin:2003qz}, and 
on eqs.~(\ref{eq:dAdx}), (\ref{eq:dBxydx}) in the Appendix 
of the present paper. 

For the 3-loop non-QCD part, the situation is more subtle, because in the 2-loop 
non-QCD contribution of eq.~(\ref{eq:Delta2M2hnonQCD}) 
we made the substitution $s = h$, implicitly dropping
3-loop order corrections of order $y_t^6 t$, 
formally of the same order as in eq.~(\ref{eq:M2hpole3c}).
However, the approximation
for the 3-loop contribution above is still justified if the 
renormalization scale $Q$ is chosen within an appropriate range. 
To see this, note that if 
$Q$ is chosen to the particular
value such that $s=h$, then the numerical error made by using $s=h$ in the 2-loop part
will vanish exactly. 
More formally, since we are interested in the 3-loop 
contributions in the limits
$s/t \ll 1$ and $y_t \gg \lambda, g, g'$, 
note that from eqs.~(\ref{eq:M2hpole}) and (\ref{eq:Delta1M2h}) we have
\beq
s = h - \frac{1}{16\pi^2} 12 y_t^2 t \lnbar(t) + \ldots 
\eeq
where the ellipses represent electroweak terms and terms suppressed by $s/t$.
Thus we see that the neglected 3-loop order terms that are of order $y_t^6 t$ will 
vanish when $Q$ is chosen so that $\lnbar(t) = 0$, 
and are correspondingly suppressed for small $\lnbar(t)$. 
In practice, the conditions $s=h$ and $\lnbar(t)=0$ imply values 
of $Q$ that are not very far apart from each other, and therefore 
this range of $Q$ is
preferred when including the 3-loop contributions above.
As we will see below, the numerical renormalization scale 
dependence of the computed $M_h$ 
is mild for a larger
range of $Q$.

%%%%%%%%%%%%%%%%%%%%%%%%%%%%%%%%%%%%%%%%%%%%%%%%%%%%%%%%%%%%%%%%
\section{Computer code implementation and numerical results\label{sec:num}}
\setcounter{equation}{0}
\setcounter{figure}{0}
\setcounter{table}{0}
\setcounter{footnote}{1}

We have implemented the Higgs pole mass calculations described 
above in a computer code library of utilities written in C, called SMH 
(for ``Standard Model Higgs"). The code can be downloaded from the authors' 
web pages \cite{webpages}.

The SMH program requires the use of the program TSIL (Two-loop 
Self-energy Integral Library) \cite{TSIL}, which is used to
handle the loop integrations. The 1-loop basis integrals
are evaluated in terms of logarithms, and the last 29
of the 2-loop integrals in the list eq.~(\ref{eq:I2list}) 
[starting with $S(0,t,W)$] 
are computed analytically in terms of polylogarithms by TSIL,
using formulas obtained in \cite{Broadhurst:1987ei,Scharf:1993ds,Ford:1992mv,Davydychev:1992mt,Davydychev:1993pg,Berends:1994ed,Caffo:1998du,Martin:2003qz}. 
The other 38 integrals are computed numerically by TSIL; this requires 
only 12 calls of the function {\tt TSIL\_Evaluate}.
The program SMH is distributed with a file {\tt README.txt}, 
which gives complete instructions for building and using 
it, as well as several example and test programs. Most user applications, like 
the example programs provided, will make use of a static archive called 
{\tt libsmh.a}, which can be linked to by C or C++ programs.

The functionality implemented in SMH includes the following: 
\begin{itemize}
%%%%%%%%%%%%%%%%%%%%%%%%%
\item {\tt SMH\_RGrun} performs the renormalization group running of 
$\lambda, y_t, g_3, g, g', m^2, v$ at up to 3-loop order, using the
$\MSbar$ beta functions and Higgs anomalous dimension 
given in refs.~\cite{MVI,MVII,Jack:1984vj,MVIII}, \cite{FJJ}, 
\cite{Chetyrkin:2013wya,Bednyakov:2013eba}. 
(At this writing, the lighter fermion
Yukawa couplings $y_b, y_\tau, y_c$ are not included, but they will be 
in a future release, as an option.)
%%%%%%%%%%%%%%%%%%%%%%%%%
\item {\tt SMH\_Find\_vev} and {\tt SMH\_Find\_m2} implement the minimization
of the Landau gauge effective potential for the Standard Model,
at up to 2-loop order \cite{FJJ} with leading 
3-loop corrections \cite{Martin:2013gka}, using 
eqs.~(4.18)-(4.21)
of ref.~\cite{Martin:2014bca}. The function {\tt SMH\_Find\_vev} finds $v$, given
$m^2, \lambda, y_t, g, g', g_3$ at a renormalization scale $Q$, 
while the function {\tt SMH\_Find\_m2} does the inverse task of finding
$m^2$, given $v, \lambda, y_t, g, g', g_3$ at $Q$. 
%%%%%%%%%%%%%%%%%%%%%%%%%
\item {\tt SMH\_Find\_Mh} and {\tt SMH\_Find\_lambda} implement the 2-loop Higgs pole mass
of eqs.~(\ref{eq:M2hpole}) and 
(\ref{eq:Delta1M2h})-(\ref{eq:Delta2M2hnonQCD}), with the leading 3-loop 
corrections from eqs.~(\ref{eq:M2hpole3a})-(\ref{eq:M2hpole3c}). 
The function {\tt SMH\_Find\_Mh}
finds $M_h$ given
$\lambda, v, y_t, g, g', g_3$ at $Q$, while the function {\tt SMH\_Find\_lambda} 
does the inverse, finding $\lambda$
given $M_h$ and $v, y_t, g, g', g_3$ at $Q$.
%%%%%%%%%%%%%%%%%%%%%%%%%
\end{itemize}
The user can choose various different loop-order approximations, as illustrated 
in the examples below, with the default being to use 
the complete set of available corrections.
Stand-alone command-line programs corresponding to each of the above
library functions are also included in the SMH package. 
We also include example programs that produce the data for the figures below.
We plan to maintain and improve the SMH code indefinitely,
and welcome bug reports or suggestions.

For purposes of illustration, consider as benchmark inputs 
[taken from ref.~\cite{Buttazzo:2013uya} version 2, eqs.~(55)-(59)]:
\beq
m^2(M_t) &=& -\mbox{(93.36 GeV})^2,
\label{eq:inputm2}
\\
\lambda(M_t) &=& 0.12711,
\label{eq:inputlambda}
\\
y_t(M_t) &=& 0.93558,
\label{eq:inputyt}
\\
g_3(M_t) &=& 1.1666,
\\
g(M_t) &=& 0.64822,
\\
g'(M_t) &=& 0.35761.
\label{eq:inputgp}
\eeq
where $Q = M_t = 173.10$ GeV is the input scale.
From these, we find our benchmark value
by minimizing the effective potential with
leading 3-loop corrections:
\beq
v(M_t) &=& \mbox{247.039 GeV}.
%247.03931235397684
\label{eq:inputvev}
\eeq
If only the full 2-loop corrections were included, the result
would be $v(M_t) = 247.381$ GeV.
%247.38110804098838

\begin{figure}[!t]
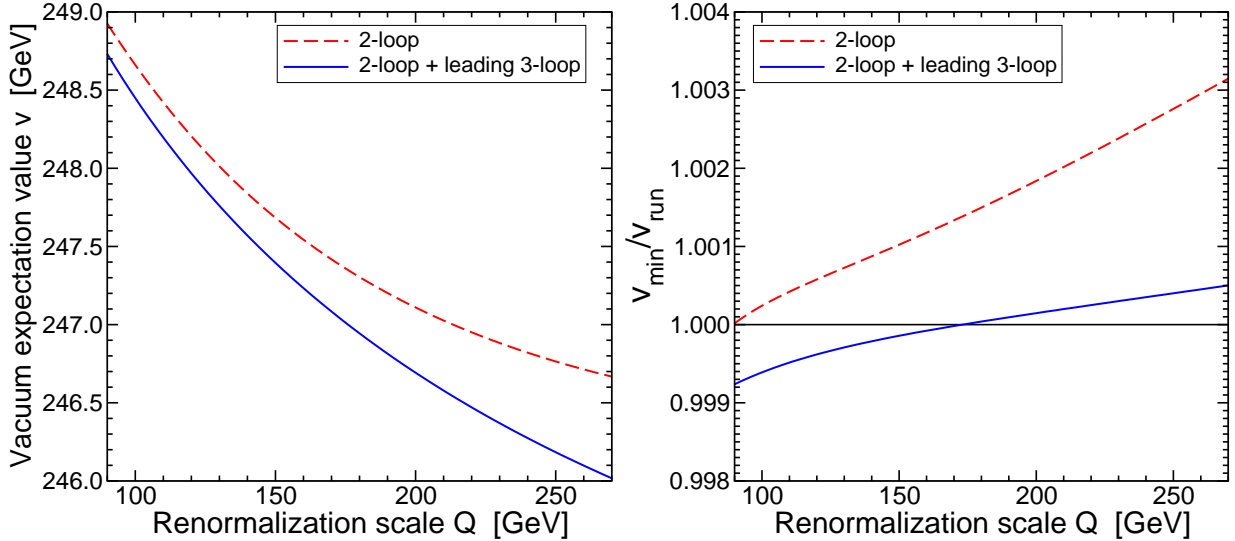

\begin{minipage}[]{0.49\linewidth}
\includegraphics[width=\linewidth,angle=0]{vev.eps}
\end{minipage}
\begin{minipage}[]{0.49\linewidth}
\includegraphics[width=\linewidth,angle=0]{vevratio.eps}
\end{minipage}
\begin{minipage}[]{0.95\linewidth}
\caption{The Standard Model Higgs VEV, $v(Q)$, obtained from minimization 
of the effective potential is shown in the left panel as a function of the renormalization scale $Q$.
The dashed line shows the results for the 2-loop
minimization condition of eqs.~(4.18)-(4.20) of ref.~\cite{Martin:2014bca},
while the solid line is the 2-loop plus leading 3-loop result obtained by including
also eq.~(4.21) of the same reference.
The input parameters
$m^2, \lambda, y_t, g_3, g, g'$ are 
obtained at the scale $Q$ by 3-loop renormalization group running 
starting from eqs.~(\ref{eq:inputm2})-(\ref{eq:inputgp}).
The right panel shows the ratio of $v(Q)$ to
the value $v_{\rm run}(Q)$  obtained from directly running it
using its renormalization group equation and input value
eq.~(\ref{eq:inputvev}). 
\label{fig:vev}}
\end{minipage}
\end{figure}

The variation of $v(Q)$ with $Q$ is shown in Figure \ref{fig:vev}. 
To make the figure,
the input parameters $m^2, \lambda, y_t, g_3, g, g'$ were run from the 
input scale to $Q$ using 3-loop renormalization group equations. 
In the left panel of Figure \ref{fig:vev}, we show the results for the 2-loop
minimization condition of eqs.~(4.18)-(4.20) of 
ref.~\cite{Martin:2014bca} as the dashed line,
while the solid line is the 2-loop plus leading 3-loop result obtained by including
also eq.~(4.21) of the same reference. The right panel shows the ratio of $v(Q)$ to
the value $v_{\rm run}(Q)$  obtained from directly running it
using its renormalization group equation and input value
eq.~(\ref{eq:inputvev}). 
%%%%%%%%%%%%%%%%%%%%%%%%%%%%%%%%%%%%%%%%%%%%%%%%%%%%%%%%%%%%%%
%%%%%%%%%%%%%%%%%%%%%%%%%%%%%%%%%%%%%%%%%%%%%%
The deviation of this ratio from unity is due to higher order-effects; it is seen to be less than 0.1\% for the calculation that includes the leading 3-loop effects.

\begin{figure}[!t]
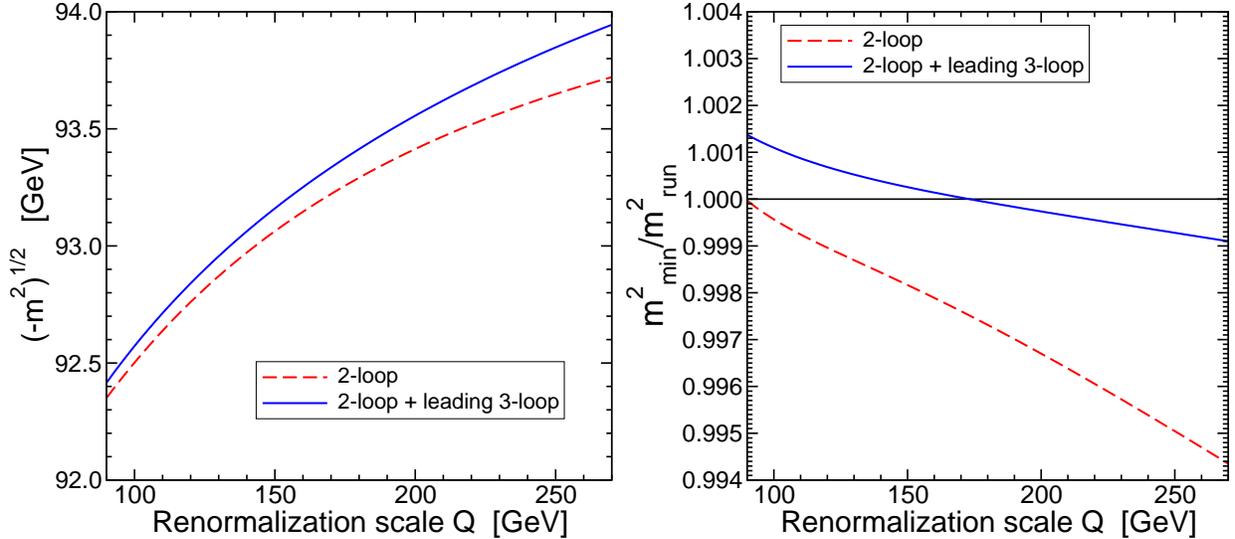

\begin{minipage}[]{0.49\linewidth}
\includegraphics[width=\linewidth,angle=0]{m2.eps}
\end{minipage}
\begin{minipage}[]{0.49\linewidth}
\includegraphics[width=\linewidth,angle=0]{m2ratio.eps}
\end{minipage}
\begin{minipage}[]{0.95\linewidth}
\caption{The Standard Model Lagrangian Higgs squared mass parameter, 
obtained from minimization 
of the effective potential is shown in the left panel as $\sqrt{-m^2}$ as a function of the renormalization scale $Q$.
The dashed line shows the results for the 2-loop
minimization condition of eqs.~(4.18)-(4.20) of ref.~\cite{Martin:2014bca},
while the solid line is the 2-loop plus leading 3-loop result obtained by including
also eq.~(4.21) of the same reference.
The input parameters
$v, \lambda, y_t, g_3, g, g'$ are 
obtained at the scale $Q$ by 3-loop renormalization group running 
starting from eqs.~(\ref{eq:inputlambda})-(\ref{eq:inputvev}).
The right panel shows the ratio of $m^2(Q)$ to
the value $m^2_{\rm run}(Q)$  obtained from directly running it
using its renormalization group equation and input value
eq.~(\ref{eq:inputm2}). 
\label{fig:m2}}
\end{minipage}
\end{figure}

In Figure \ref{fig:m2}, we reverse the roles of $m^2$ and $v$, by showing 
the dependence of the Higgs Lagrangian mass parameter
$m^2(Q)$ obtained by minimizing the effective potential, this time with the VEV $v(Q)$
as an input parameter. To make the figure,
the input parameters $v, \lambda, y_t, g_3, g, g'$ were run from the 
input scale to $Q$ using 3-loop renormalization group equations. 
In the left panel of Figure \ref{fig:m2}, we show $\sqrt{-m^2}$ obtained from 
the 2-loop
minimization condition of eqs.~(4.18)-(4.20) of 
ref.~\cite{Martin:2014bca} as the dashed line,
while the solid line is the 2-loop plus leading 3-loop result obtained by including
also eq.~(4.21) of the same reference.
The right panel shows the ratio of $m^2(Q)$ to
the value $m^2_{\rm run}(Q)$  obtained from directly running it
using its renormalization group equation and input value
eq.~(\ref{eq:inputm2}). 
%%%%%%%%%%%%%%%%%%%%%%%%%%%%%%%%%%%%%%%%%%%%%%%%%%%%%%%%%%%%%%
%%%%%%%%%%%%%%%%%%%%%%%%%%%%%%%%%%%%%%%%%%%%%%

In Figure \ref{fig:MhQ}, we show
results for the Higgs pole mass $M_h$ as a function of the renormalization scale $Q$.
To make the figure,
the input parameters $\lambda, y_t, g, g', g_3, v$ were run from the 
input scale to $Q$ using 3-loop renormalization group equations. 
The lower solid (blue) line is the 2-loop calculation of 
eqs.~(\ref{eq:M2hpole}) and
(\ref{eq:Delta1M2h})-(\ref{eq:Delta2M2hnonQCD}),
while the upper solid (black) line
includes also the leading 3-loop contributions of 
eqs.~(\ref{eq:M2hpole3a})-(\ref{eq:M2hpole3c}). 
The results at the input scale $Q=173.1$ GeV are
$M_h = 125.789$ GeV and $M_h = 125.818$, 
%$M_h = 125.789030292594319$ GeV and $M_h = 125.817674445041746$, 
respectively.
We also show 
the tree-level approximation $\sqrt{2 \lambda} v$ 
as the dotted line, and 
the 1-loop approximation
obtained from eqs.~(\ref{eq:M2hpole}) and (\ref{eq:Delta1M2h})
as the short-dashed line, and the 1-loop approximation with
the 2-loop QCD corrections from (\ref{eq:Delta2M2hQCD}) 
included as the long-dashed line.

Figure \ref{fig:MhQcloseup} is a close-up of the previous figure, 
to illustrate the scale dependence
more clearly for the full 2-loop and leading 3-loop approximations. 
%%%%%%%%%%%%%%%%%%%%%%%%%%%%%%%%%%%%%%%%%%%%%%%%%%%%%%%%%%%%%%
\begin{figure}[!t]
\vspace{-0.9cm}
\begin{minipage}[]{0.62\linewidth}
\includegraphics[width=\linewidth,angle=0]{MhQ.eps}
\end{minipage}
\begin{minipage}[]{\linewidth}
\caption{The calculated Higgs pole mass $M_h$ as a function
of the renormalization scale $Q$, in various approximations.
The input data at $Q$ are obtained from
3-loop renormalization group running 
of $\lambda, y_t, g, g', g_3, v$ starting from 
eqs.~(\ref{eq:inputlambda})-(\ref{eq:inputvev}).
The dotted (green) 
line is the tree-level approximation $\sqrt{2 \lambda} v$. 
The short-dashed (orange)
line is the 1-loop approximation
obtained from eqs.~(\ref{eq:M2hpole}) and (\ref{eq:Delta1M2h}).
The long-dashed (red) line is the 1-loop approximation with
the 2-loop QCD corrections from eq.~(\ref{eq:Delta2M2hQCD}).
The lower solid (blue) line is the 2-loop $M_h$ 
as calculated from
eqs.~(\ref{eq:M2hpole}) and
(\ref{eq:Delta1M2h})-(\ref{eq:Delta2M2hnonQCD}), while the
upper solid (black) line
also includes the leading 3-loop corrections of 
eqs.~(\ref{eq:M2hpole3a})-(\ref{eq:M2hpole3c}). 
\label{fig:MhQ}}
\end{minipage}
%\end{figure}
%%%%%%%%%%%%%%%%%%%%%%%%%%%%%%%%%%%%%%%%%%%%%%%%%%%%%%%%%%%%%%
%%%%%%%%%%%%%%%%%%%%%%%%%%%%%%%%%%%%%%%%%%%%%%%%%%%%%%%%%%%%%%
%\begin{figure}[!t]
\begin{minipage}[]{0.63\linewidth}
\vspace{0.3cm}
\begin{flushright}
\includegraphics[width=\linewidth,angle=0]{MhQcloseup.eps}
\end{flushright}
\end{minipage}
\begin{minipage}[]{\linewidth}
\caption{A close-up of the dependence of the calculated $M_h$ on $Q$, as in Fig.~\ref{fig:MhQ}.
The lower (blue) line is the full 2-loop $M_h$ 
as calculated from
eqs.~(\ref{eq:M2hpole}) and
(\ref{eq:Delta1M2h})-(\ref{eq:Delta2M2hnonQCD}). 
The upper (magenta) line is the full 2-loop plus the 3-loop QCD contribution
of eqs.~(\ref{eq:M2hpole3a})-(\ref{eq:M2hpole3b}), not shown in Fig.~\ref{fig:MhQ}. The
middle (black) line is the full 2-loop plus the 3-loop corrections of 
eqs.~(\ref{eq:M2hpole3a})-(\ref{eq:M2hpole3c}),
with the left dot marking the case $s=h$ and the right dot
marking the case $\lnbar(t)=0$.  
\label{fig:MhQcloseup}}
\end{minipage}
\end{figure}
%%%%%%%%%%%%%%%%%%%%%%%%%%%%%%%%%%%%%%%%%%%%%%%%%%%%%%%%%%%%%%
The lower (blue) line is again the full 2-loop $M_h$ 
as calculated from eqs.~(\ref{eq:M2hpole}) and
(\ref{eq:Delta1M2h})-(\ref{eq:Delta2M2hnonQCD}).
For comparison, we also show the result for the full 2-loop plus the 3-loop QCD 
contribution of eqs.~(\ref{eq:M2hpole3a})-(\ref{eq:M2hpole3b}), without including
the non-QCD 3-loop corrections, as the upper (magenta) line. This has a much stronger
scale dependence than the 2-loop result, despite the formal independence of $M_h$
with respect to $Q$ through terms of 3-loop order involving $g_3$. Including the
non-QCD $y_t^6 t$ contributions from eq.~(\ref{eq:M2hpole3c}) 
yields the middle (black) line, which again
has a mild scale dependence comparable to the 2-loop result. The residual scale 
dependence is due to higher order effects. Note that, as can be seen by comparing 
eqs.~(\ref{eq:M2hpole3b}) and (\ref{eq:M2hpole3c}), the 3-loop QCD and 3-loop $y_t^6 t$ 
contributions contribute with opposite sign, and have an opposite scale dependence.

The points with $s=h$ and $\lnbar(t)=0$ are marked with dots on the leading
3-loop $M_h$ line in Figure \ref{fig:MhQcloseup}. As argued in the previous section,
the range of $Q$ near these points is preferred due to the treatment of the 
2-loop corrections. In particular, the choice  of $Q$ that makes 
$\lnbar(t)=0$ is easy to implement as a natural standard. Given the value of
the running top-quark mass, and the observed mild scale dependence 
in this region, a fixed value of, say, $Q=160$ GeV would also make sense.

\begin{figure}[!t]
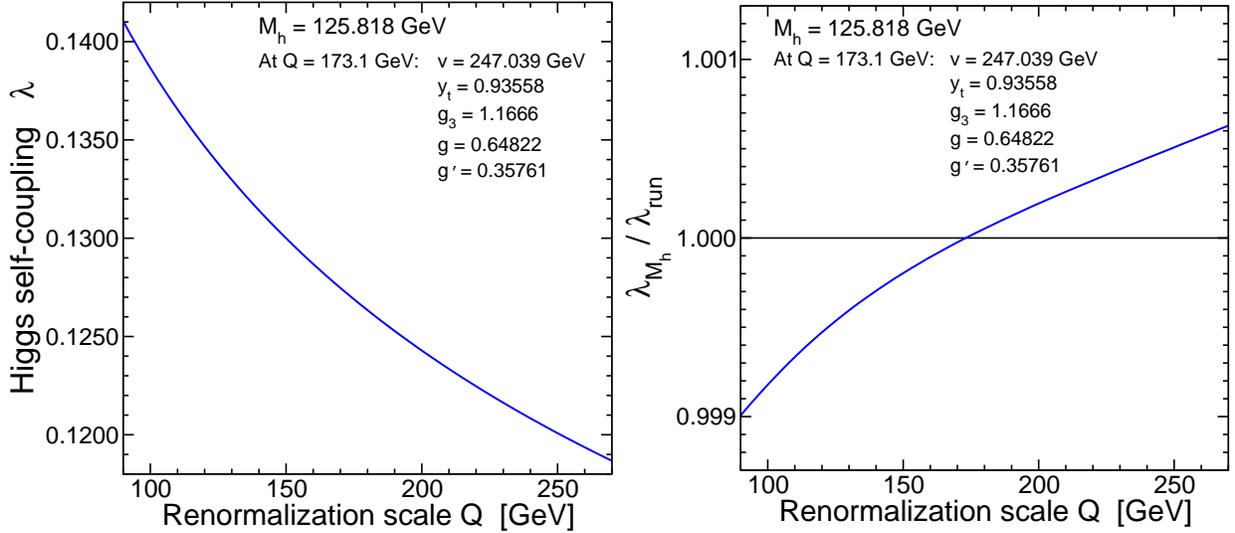

\begin{minipage}[]{0.49\linewidth}
\includegraphics[width=\linewidth,angle=0]{kQ.eps}
\end{minipage}
\begin{minipage}[]{0.49\linewidth}
\includegraphics[width=\linewidth,angle=0]{kratioQ.eps}
\end{minipage}
\begin{minipage}[]{0.95\linewidth}
\caption{The Higgs self-coupling parameter, $\lambda_{M_h}(Q)$ as calculated from
a fixed pole mass $M_h = 125.818$ GeV using
eqs.~(\ref{eq:M2hpole}) and
(\ref{eq:Delta1M2h})-(\ref{eq:Delta2M2hnonQCD}) and 
eqs.~(\ref{eq:M2hpole3a})-(\ref{eq:M2hpole3c}), with
$y_t, g, g', g_3, v$ obtained at the scale $Q$ by 3-loop renormalization group running 
starting from eqs.~(\ref{eq:inputyt})-(\ref{eq:inputvev}).
The right panel shows the ratio of $\lambda_{M_h}(Q)$ to
the value $\lambda_{\rm run}(Q)$  obtained from directly running it
using its renormalization group equation and input value
eq.~(\ref{eq:inputlambda}). 
\label{fig:lambda}}
\end{minipage}
\end{figure}

In the left panel of Figure \ref{fig:lambda}, we show the scale dependence
of $\lambda(Q)$ obtained from eqs.~(\ref{eq:M2hpole}) and
(\ref{eq:Delta1M2h})-(\ref{eq:Delta2M2hnonQCD}) and 
eqs.~(\ref{eq:M2hpole3a})-(\ref{eq:M2hpole3c}), 
with the same input parameters $v, y_t, g, g', g_3$ at $Q = 173.1$ GeV,
but now using a fixed pole mass $M_h = 125.818$ GeV 
as the input. This value is chosen so that the calculated Higgs self-coupling at the 
input scale agrees with eq.~(\ref{eq:inputlambda}).
In the right panel, we show
the ratio of $\lambda_{M_h}(Q)$ determined in this way to 
$\lambda_{\rm run}(Q)$ obtained by directly running it from the input value
eq.~(\ref{eq:inputlambda})
using its 3-loop renormalization group equation. 
%%%%%%%%%%%%%%%%%%%%%%%%%%%%%%%%%%%%%%%%%%%%%%%%%%%%%%%%%%%%%%
%%%%%%%%%%%%%%%%%%%%%%%%%%%%%%%%%%%%%%%%%%%%%%
As expected, the ratio is very close
to 1 for all values of $Q$; the two versions of $\lambda$ 
would be visually indistinguishable in the left panel.
These results illustrate the renormalization group scale 
independence through 2-loop and 3-loop QCD order that we verified
analytically as described above, with small discrepancies less than 0.1\% 
coming from 3-loop $y_t^6 t$ and from sub-leading 3-loop and
higher-order effects.

\section{Outlook\label{sec:outlook}}
\setcounter{equation}{0}
\setcounter{figure}{0}
\setcounter{table}{0}
\setcounter{footnote}{1}
\vspace{0.25cm}

In this paper, we have obtained the pole mass of the Higgs boson, $M_h$, 
including full 2-loop and leading 3-loop corrections, 
in the $\MSbar$ scheme. The 
calculation was done in Landau gauge, in order to match with existing multi-loop
calculations of the effective potential used to eliminate $m^2$ by relating it to the 
VEV $v$ and the other Lagrangian parameters. 
The inputs to the calculation 
are the $\MSbar$ running parameters of the theory, $v, \lambda, y_t, g, 
g', g_3$. Other observables, such as the pole masses of the top quark 
and the $W,Z$ bosons, are not inputs to the calculation, and are to be 
calculated separately. A possible advantage to this strategy is that 
future refinements in calculations and measurements 
of those other observable quantities 
will not be entangled with the calculation of the Higgs pole mass. 
Previous results for the 2-loop corrections 
\cite{Bezrukov:2012sa,Degrassi:2012ry,Buttazzo:2013uya} to the Higgs 
mass were organized in a different way, and in the case of the non-QCD 
corrections \cite{Degrassi:2012ry,Buttazzo:2013uya} were given only in 
the form of simple interpolating formulas, making comparison with the 
present paper not practical. Our full analytic results are contained in 
an ancillary electronic file, and a computer code called SMH is 
provided \cite{webpages}, implementing the results for $M_h$, the effective potential 
minimization, and renormalization group running.

Because there is no way of directly measuring the Higgs self-coupling 
parameter accurately in the immediate future, the measurement of the Higgs 
mass is the best way to determine $\lambda$, assuming the validity of 
the Standard Model, with variations related approximately by 
\beq 
\Delta\lambda &=& 0.00205 (\Delta M_h/\mbox{GeV}) . 
\eeq 
From the renormalization scale variation and the 
magnitudes of the leading 3-loop QCD and non-QCD effects, 
we make a very rough estimate of the theoretical 
uncertainty on $M_h$ of 100 MeV, or about $0.1\%$, taking $\MSbar$ quantities as 
the inputs. This does not include the effects of reducible parametric error, 
notably the dependence on the uncertainties in the top-quark Yukawa 
coupling (or mass) and the QCD coupling. The future experimental error 
in $M_h$ has been estimated \cite{Dawson:2013bba} to be perhaps 100 MeV 
(50 MeV) with 300 fb$^{-1}$ (respectively 3000 fb$^{-1}$) at the LHC, 
and of order 30 MeV or less at future $e^+e^-$ colliders. We conclude 
that more refined 3-loop order and quite possibly 4-loop order 
corrections to $M_h$ will be necessary in order to 
make the theoretical error small compared to the foreseeable 
experimental error, discounting the parametric uncertainties that may be 
reducible by independent calculations and measurements. At the least, a further
refinement of the 3-loop $M_h$ calculation would serve to firm up an estimate of the 
theoretical error.

Besides applications within the Standard Model, the result may find use 
in extensions of the Standard Model, including supersymmetry. The most 
straightforward interpretation of the current LHC searches for 
supersymmetry is that the superpartners, if they exist, are sufficiently 
heavy that the Standard Model can be treated as an effective theory with 
other new physics nearly decoupled. The direct observation that the 
Higgs mass is relatively large compared to most pre-LHC expectations 
within supersymmetry can be taken as indirect evidence of the same 
thing. In the past, many attempts to compute the Higgs mass within 
supersymmetry have calculated directly within the full softly broken 
supersymmetric theory in the Feynman diagrammatic 
\cite{Haber:1990aw}-\cite{Hahn:2013ria} and effective potential 
approximation \cite{Carena:1995wu}-\cite{Martin:2002wn} approaches. 
However, it now seems to us that with very heavy superpartners, the 
effective field theory and renormalization group resummation strategy 
\cite{Haber:1993an}-\cite{Draper:2013oza} for calculating the Higgs mass 
is probably the best one. One can match the supersymmetric theory onto the 
Standard Model parameters as an effective theory at some scale or scales 
comparable to the most important superpartner masses (probably the top 
squarks), and then run the parameters of the theory down to a scale 
comparable to $M_t$, and there compute $M_h$ within the Standard Model. 
In that case, the results obtained here may be a useful ingredient.

\section*{Appendix: Some loop integral identities}\label{sec:appendix}
\renewcommand{\theequation}{A.\arabic{equation}}
\setcounter{equation}{0}
\setcounter{footnote}{1}

This Appendix contains some loop integral identities that are useful for 
processing and simplifying the 2-loop Higgs pole mass. Other useful identities
in the notation of the present paper can be found in 
refs.~\cite{Martin:2003qz,TSIL,Martin:2003it}

First, the derivatives of 1-loop basis functions,
obtained by dimensional analysis and
integration by parts, are:
\beq
\frac{\partial}{\partial x}{\bf A}(x) &=& (d/2-1) {\bf A}(x)/x, 
\\
\frac{\partial}{\partial x}{\bf B}(x,y) &=&
\Bigl [ (d-3) (x-y-s) {\bf B}(x,y) + (d-2) \lbrace
(x+y-s){\bf A}(x)/2x\,
- {\bf A}(y)
\rbrace \Bigr ]/\Delta_{sxy} ,
\nonumber \\ &&
\eeq
where $d = 4 - 2 \epsilon$ is the number of 
spacetime dimensions and
$\Delta_{abc} \equiv a^2 + b^2 + c^2 - 2 a b - 2 a c - 2 b c$.
Using the expansions for small $\epsilon$,
\beq
{\bf A}(x) &=& -x/\epsilon + A(x) + \epsilon A_\epsilon(x) 
+ {\cal O}(\epsilon^2),
\\
{\bf B}(x,y) &=& 1/\epsilon + B(x,y) + \epsilon B_\epsilon(x,y) 
+ {\cal O}(\epsilon^2),
\eeq
one then obtains
\beq
\frac{\partial}{\partial x} A(x) &=& A(x)/x + 1 \>=\> \lnbar(x)
,
\label{eq:dAdx}
\\
\frac{\partial}{\partial x} B(x,y) &=&
\{(x-y-s) [ B(x,y) - 1 ] +(x+y-s) A(x) /x - 2 A(y)] \}/\Delta_{sxy},
\label{eq:dBxydx}
\eeq
and the expansions for small $G$:
\beq
A(G) &=& G\, \lnbar(G) - G,
\label{eq:AG}
\\
B(0,G) &=& B(0,0) + G [3 - B(0,0) - \lnbar(G)]/s + {\cal O}(G^2),
\\ 
B(G,G) &=& B(0,0) + 2 G [3 - B(0,0) - \lnbar(G)]/s + {\cal O}(G^2),
\\ 
B(G,x) &=& B(0,x)  + {G} [3s-x-(s+x) B(0,x)
- 2 A(x) ]/{(x-s)^2}\, 
\nonumber \\ &&
+ {G \lnbar(G)}/{(x-s)}\,+ {\cal O}(G^2),
\phantom{xxxx}
\\
A_\epsilon(G) &=& G [-1 - \pi^2/12 + \lnbar(G) - \frac{1}{2} \lnbar^2(G) ]
,
\\
B_\epsilon (0,G) &=& B_\epsilon (0,0) + G[-A_\epsilon(G)/G + 2 B(0,0) 
- B_\epsilon(0,0)]/s + {\cal O}(G^2)
,
\\
B_\epsilon (G,G) &=& B_\epsilon (0,0) + 2 G[-A_\epsilon(G)/G + 2 B(0,0) 
- B_\epsilon(0,0)]/s + {\cal O}(G^2)
,
\\
B_\epsilon (G,x) &=& B_\epsilon (0,x) 
+ G \Bigl [2 A(x) - 2 A_\epsilon(x) 
+ (s+x) \bigl \{2 B(0,x)
- B_\epsilon(0,x) \bigr \} \Bigr]/{(x-s)^2} \phantom{xxxx}
\nonumber \\ && 
+ {A_\epsilon(G)}/{(x-s)} + {\cal O}(G^2). 
\label{eq:BepsGx}
\eeq

Some identities between basis integrals that hold for non-generic
squared mass arguments are the threshold identities:
\beq
&& \lim_{s\rightarrow x} B(0,x) \>=\> 1 - A(x)/x,
\label{eq:B0h}
\\
&& \lim_{s\rightarrow x} \left [
\overline{T}(0,0,x) + T(x,0,0) \right ] \>=\> -1,
\label{eq:Th00}
\eeq
and the general relations
\beq
I(0,0,x) &=& A(x) - A(x)^2/2x\, -x (1 + \pi^2/6) 
,
\label{eq:I00x}
\\
I(0,x,x) &=& 2 A(x) - A(x)^2/x \,- 2 x,
\\
{\overline T}(0,0,x) &=& -T(x,0,0)
+ [-s + 2 A(x) - A(x)^2/x + (s+x) B(0,x) 
\nonumber \\ &&
- (s+x) A(x) B(0,x)/x - s B(0,x)^2]/(s-x),
\\
{\overline T}(0,0,0) &=& -[B(0,0) -1]^2/2,
\\
U(0,x,0,0) &=& (1-x/s) T(x,0,0) + B(0,0) B(0,x) + A(x)  B(0,0)/x 
\nonumber \\ &&
+(1-x/s) B(0,x) - I(0,0,x)/s + 2 - x/s.
\label{eq:U0x00}
\eeq
Other identities of similar type that express redundancies among the 
basis integrals for non-generic squared mass arguments 
and were used here have appeared as eqs.~(A.14), (A.15),
and (A.17)-(A.20) of 
ref.~\cite{Martin:2003it}.

\vspace{0.2cm}

\noindent {\it Acknowledgments:} 
This work was supported in part by the National
Science Foundation grant numbers PHY-1068369 and PHY-1417028. 
This work was supported in part by the 
National Science Foundation under Grant No. PHYS-1066293 
and the hospitality of the Aspen Center for Physics. 
DGR was supported in part by a grant from the Ohio Supercomputer Center.

%%%%%%%%%%%%%%%%%%%%%%%%%%%%%%%%%%%%%%%%%%%%%%%%%%%%%%%%%%%%%%%%%%%%%%%%%%%%

%%%%%%%%%%%%%%%%%%%%%%%%%%%%%%
\end{document}